# Modeling crater formation in femtosecond-pulse laser damage from basic principles

Robert A. Mitchell,* Douglass W. Schumacher, and Enam A. Chowdhury

*Department of Physics, The Ohio State University, 191 W. Woodruff Ave, Columbus, OH 43210*
*Corresponding author: mitchell.911@osu.edu*

We present the first fundamental simulation method for the determination of crater morphology due to femtosecond-pulse laser damage. To this end we have adapted the particle-in-cell (PIC) method commonly used in plasma physics for use in the study of laser damage, and developed the first implementation of a pair-potential for PIC codes. We find that the PIC method is a complementary approach to modeling laser damage, bridging the gap between fully ab-initio molecular dynamics approaches and empirical models. We demonstrate our method by modeling a femtosecond-pulse laser incident on a flat copper slab, for a range of intensities.

Laser damage from femtosecond pulses is known for its precise and predictable nature. The lack of collateral damage which normally accompanies longer pulses makes femtosecond pulse lasers exceptional tools for precise micromachining and materials characterization.[1] On the other hand, the generation of the intense laser pulses themselves requires the avoidance of laser damage during amplification and compression. A better understanding of the fundamental processes behind laser damage could improve the use of short pulse laser material modification and characterization, and help to guide improvements in damage-resistant optics.

Many theoretical tools have been developed over the last several decades for the study of laser damage.[2-4] The majority of these techniques fall into one of two categories: small spatial scale, ab-initio simulation techniques such as the Molecular Dynamics method; or large spatial scale, empirical or rate-equation models. Some of these methods have met with considerable success for aspects of this problem, but a microscopic model of crater formation has not yet been demonstrated. Rate-equation or empirical model approaches are generally only used to calculate quantities like the damage threshold fluence, but each makes an assumption as to what constitutes damage. The two-temperature model, for example, assumes that damage occurs when a given amount of material at the surface reaches the melting temperature of the material.[3] An ionization-based rate-equation approach assumes that damage occurs when the electron density reaches a pre-defined threshold, typically the plasma critical density.[4] Molecular Dynamics approaches, on the other hand, provide a powerful view of the atomic dynamics, but the limited spatial extent of these computationally-intensive simulations prevents modeling of crater morphology, instead focusing on such things as damage threshold measures and ablation rates.

Both of these approaches have a disconnect from experiment. For the experimentalist, *crater morphology is the primary observable.* Laser damage experiments generally measure a damage crater, either simply its appearance or its width or depth, and usually after multiple laser pulses. To compare such an experiment to existing models, the experimental results have to be interpolated from multi-shot to single-shot results, and then further interpolated from, for example, a width or depth measurement to the threshold fluence.[5] Currently, there are no simulation methods able to treat crater morphology in a fundamental way so as to compare directly to experiment.

Previously, we presented a proof-of-concept that the particle-in-cell (PIC) simulation method could provide a new and complimentary technique for modeling damage, and in particular crater formation allowing for a direct comparison to experiment in both single-shot and few-shot studies.[6] PIC simulations are fundamental in nature in that they directly integrate the equations of motion, such as the Maxwell and Lorentz equations, but also make enough simplifying approximations that modeling the full laser focus and crater formation is possible. The two primary approximations of a PIC code are the statistical sampling of the target particles through the use of so-called macroparticles, and a partial discretization of space where particle interactions are discretized and occur via nodes although particle positions remain continuous, as illustrated in Fig. 1A. A more detailed explanation of the PIC method can be found in Ref. 6.

We present here a much improved method for modeling the atomic transport leading to crater formation with PIC codes. Unlike the previous pair-potential algorithm presented in Ref. 6, this new algorithm is stable and energy conserving for indefinitely long simulation times, and the simulation runs more than an order of magnitude faster due to more rapid numerical convergence with particle number and cell size. It also avoids the unphysical, numerical artifacts of the earlier approach.

The key challenge to using PIC to study laser damage is the addition of a model to include the effect of a pair-potential between particles so that interatomic forces can be treated but that is also consistent with the use of macroparticles and the partial discretization of space. We do this by calculating a force on each macroparticle that is derived from the PIC statistical representation of the local environment at each macroparticle's position, the details of which are outlined below, and illustrated in Fig. 1B. The PIC cycle begins by interpolating the weight (ie. number of actual particles represented) of each macroparticle to the nearest surrounding nodes in order to calculate a node density. See Fig. 1A for the geometry of PIC discretization. The density at the nodes is then used to calculate a density gradient at the half-node (cell-centered) positions. Later in the cycle, this node density and density gradient is interpolated in a self-consistent fashion back to the macroparticles, in order to calculate the local environment at each macroparticle position. The local environment, combined with an assumption on lattice structure, is enough to calculate the interatomic forces. For the purpose of simplifying the development of this approach we have assumed that the atoms form a cubic lattice. However, this method can be readily adapted for a more realistic lattice such as hexagonal close-packed.

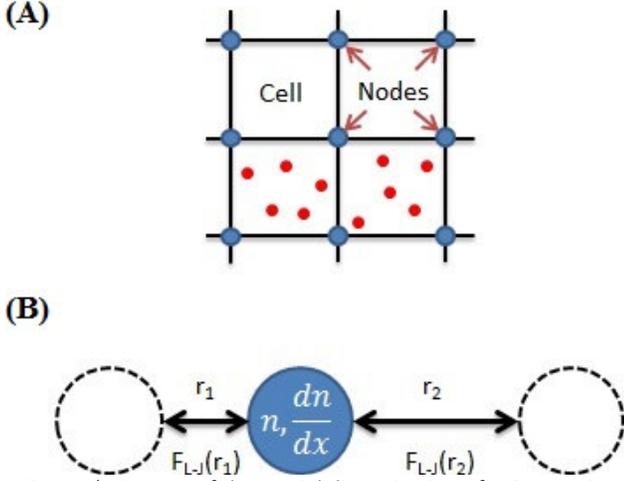

Fig. 1. A) Diagram of the partial discretization of PIC space into cells and nodes, with the particles (red circles) free to have continuous position. B) Illustration of the pair-potential calculation between an atom at a macroparticle's position and its nearest-neighbor atoms.

To determine the effective interatomic force, we calculate where the nearest-neighbor atoms would be relative to an atom located at each macroparticle position as follows. We calculate the average inter-particle distance, $\bar{r} = n^{-1/3}$, and the change in the average inter-particle distance using the local density and density gradient. For example, in the x-direction:

$$\frac{\partial \bar{r}}{\partial x} = -\frac{1}{3}\frac{\partial n}{\partial x} n^{-4/3} \qquad (1)$$

Here $n$ is the local number density. We calculate the distance to an atom's nearest-neighbors by treating the interatomic distance variation in the local vicinity of the atom as linear, so $\nabla \bar{r}$ is constant. Then, the average inter-particle distance $\bar{r}$ midway between two atoms will be the distance between those two atoms. In the x-direction, the distances to an atom's two nearest neighbors, $\Delta r_1$ and $\Delta r_2$, will be:

$$\Delta r_{1,2} = \bar{r}\left(1 \pm \frac{1}{2}\frac{\partial \bar{r}}{\partial x}\right)^{-1} \qquad (2)$$

The above algorithm takes the current PIC representation of density variation in the simulation and maps it to atomic separation distances for the calculation of interatomic forces using a pair-potential. This choice of algorithm was essential to avoid PIC instabilities and is fully compatible with any pair-potential, such as the Morse potential, Lennard-Jones (L-J) potential, or Embedded-Atom Method.[7,8] We have chosen the L-J potential for simplicity, and because it has been shown to approximate the relevant material properties reasonably well.[7] The L-J potential $U$ can be written as:

$$U(\Delta r) = D\left(\left(\frac{r_0}{\Delta r}\right)^{12} - 2\left(\frac{r_0}{\Delta r}\right)^6\right) \qquad (3)$$

Here $D$ is the dissociation energy, $r_0$ is the equilibrium distance, and $\Delta r$ is the separation distance between atoms.

Having achieved a stable PIC cycle, the spatial discretization of the PIC method still causes errors in the interatomic forces as the cell size is increased. A sufficiently large cell size that permits rapid evaluation has two effects: one, the particle density is smoothed over, effectively damping the repulsive component of the force; two, the large cell size extends the range of the attractive interaction to a cell length, which effectively increases the binding energy of atoms to the target. To counter the damping of the repulsive component of the force, we used an altered L-J potential that is infinite at distances closer than the equilibrium distance. To account for the increased attractive component of the force, a multiplicative correction factor for the force is calculated by integrating over the force applied to a single atom leaving the target, traveling out to infinity.

The resulting PIC-consistent approach to a medium with a pair-potential interaction is then complete and has no free parameters that can be tuned. Once a damage evolution simulation is initialized for target density and temperature, the PIC cycle is simply evaluated for each time step as usual. The target density should represent the structure of the initial target and the temperature profile should represent the lattice temperature after the electron and lattice temperatures equilibrate. The temperature initialization can come from any method that can treat the interaction of a laser and cold target including the two temperature model, an empirical heating profile, an experimentally measured temperature distribution, or the standard PIC method. To demonstrate the full use of PIC, we modeled a short pulse laser incident on a flat copper slab target using two sequential simulations, treating first the fs laser-target interaction and then the subsequent ns target damage using the code LSP,[9] which was modified to include the pair-potential interaction model.

Although LSP and our algorithm are fully 3D, the simulations were performed in 2D to reduce computation time. The target has width and depth, but is effectively infinitely long in the remaining dimension. The laser interaction can be visualized as having a line-focus. Note that in 2D PIC, all vector quantities including electric and magnetic fields and particle momenta are still three component vectors, however the particle positions are not allowed to evolve in the third dimension. Thus, self-consistent light propagation and interaction with the target is supported.

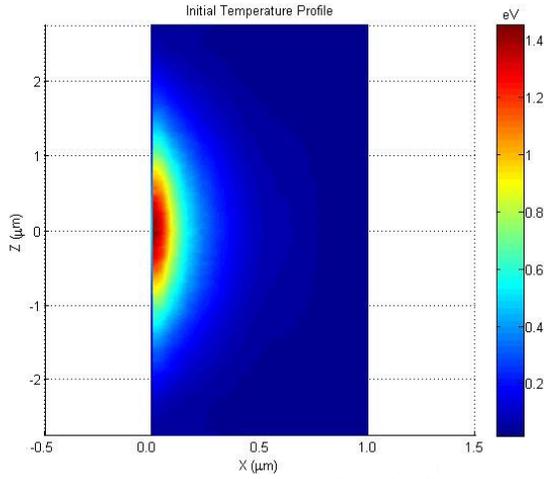

Fig. 2. Initial atomic temperature profile of a flat copper slab, as calculated via a laser-plasma interaction PIC simulation with a 2.0 J/cm$^2$, 1 μm waist, 60 fs FWHM laser pulse.

The laser modeled was a 60 fs FWHM Gaussian pulse with a central wavelength of 800 nm focused to a 1 μm waist at normal incidence for a variety of intensities. The laser-interaction simulation was done with 1/128 × 1/32 μm cell sizes in the laser propagation and transverse dimensions, respectively, with 0.0055 fs timesteps, 900 macro-electrons per cell and 25 Cu$^+$ macro-ions per cell. Values for the dissociation energy and equilibrium distance of copper atoms were taken from Ref. 10. This high-resolution simulation was allowed to run until the electrons and ions were near thermal equilibrium. The resulting lattice temperature profile used for initializing the atomic transport simulation for a 2.0 J/cm$^2$ fluence is shown in Fig. 2.

After initializing the atomic-transport simulation with the lattice temperature profile calculated from the laser-target interaction simulation, we allowed the target to evolve using our pair-potential implementation. The atomic-transport simulation was executed with significantly lower, though still numerically converged, resolution for rapid evaluation. The simulation used 10 × 10 nm cell sizes, 12 fs timesteps, and 3600 neutral Cu atoms per cell. Though we used neutral Cu atoms in this example, this method is fully compatible with using a combination of atoms, ions, and electrons as necessary, for instance, if modeling a dielectric and including ionization effects. The targets were 4.5-6 μm wide, and 0.75-1.0 μm thick, depending on the incident fluence, ensuring the targets were large enough to capture the damage morphology for these conditions. The transport simulation was allowed to evolve for several nanoseconds, until the target surface stabilized and the crater had solidified. Fig. 3 shows density profile snapshots of the target evolution during the damage process for a laser fluence of 2.0 J/cm$^2$, in units of solid density, which is 8.5 x 10$^{22}$ atoms/cm$^3$ for copper.

Fig. 3A shows the initial target geometry. As can be seen from Fig. 3B, the ablation plume is just starting to form around 40 ps. As time goes on, the plume expands, until around 1.5 ns, Fig. 3E, when the plume slowly begins to dissipate leaving a clean, well-defined damage crater after approximately 3 ns, as shown in Fig. 3H. For the purpose of illustrating the crater, Fig. 3 has a lower limit of 1% of solid density below which nothing is plotted, where in fact there is still a low density gas of Cu atoms filling the vacuum off the front surface of the target. This low density gas would take much longer to evacuate, and has no effect on the resultant damage spot, making it unnecessary to model further.

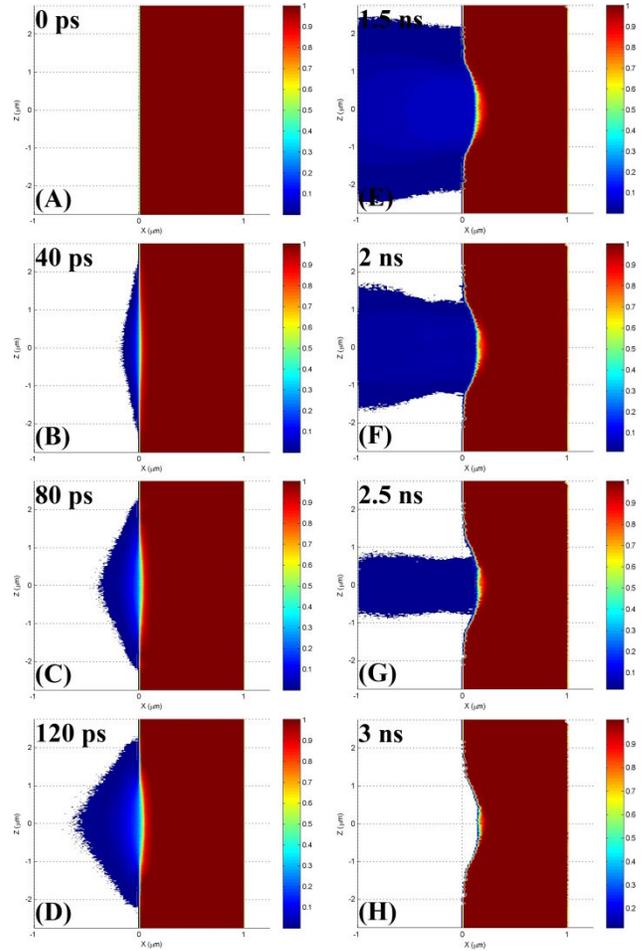

Fig. 3. Density profiles of the initial target morphology and evolution of a 2.0 J/cm$^2$ laser fluence damage crater and ablation plume in units of solid density of copper, at A) Start of simulation, B) 40 ps, C) 80 ps, D) 120 ps, E) 1.5 ns, F) 2 ns, G) 2.5 ns, H) 3 ns. Each plot has a lower density cut-off at 1% solid density.

After the crater has solidified, evolution has ceased, and the atomic-transport simulation has ended, we can measure the damage crater's width and depth to compare directly to experiment. In the case of few-shot studies, the resultant target structure, in the form of a temperature and density profile, can be used for initializing another laser-target interaction simulation followed by a transport simulation.

The resulting craters for laser fluences of 0.5 J/cm$^2$, 1.0 J/cm$^2$, and 2.0 J/cm$^2$ are shown in Figs. 4A, 4B, and 4C, respectively. To compare to recent experiments which generally used a laser that was not as tightly focused as our simulations here, we apply a procedure for extracting a threshold fluence from experimental crater morphologies to our simulation results.[11] The approach is based on interpolation using the crater width, resulting in the following relation:

$$F_{th} = F exp\left\{-\left(\frac{\Gamma}{a}\right)^2\right\} \quad (5)$$

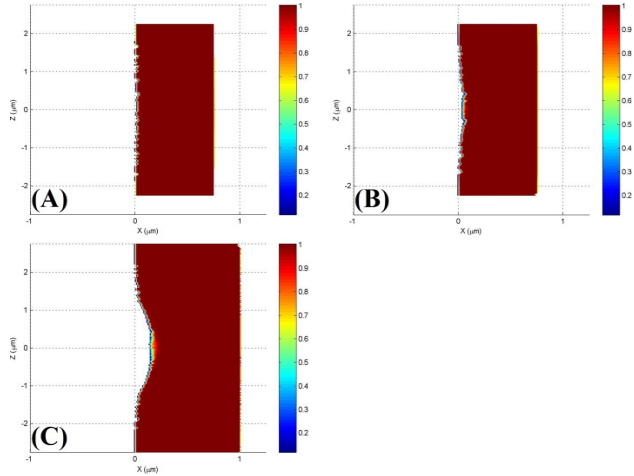

Fig. 4. Density profiles of the resultant damage crater after surface evolution has ceased, in units of solid density of copper, for A) 0.5 J/cm$^2$, B) 1.0 J/cm$^2$, and C) 2.0 J/cm$^2$. Each plot has a low density cut-off of 1% solid density.

Where $\Gamma$ is the crater diameter, $a$ is the laser beam diameter, $F$ is the incident laser fluence and $F_{th}$ is the damage threshold fluence. Using our crater and beam widths in this formula results in a damage threshold of 0.15 J/cm$^2$, which is consistent with experimental measurements of copper thresholds for the modeled laser system.[5,11]

In summary, we have presented a new simulation technique for modeling short-pulse laser damage of materials. This algorithm represents a complimentary approach to current methods, with different strengths, limitations, and capabilities. Modeling laser damage with PIC requires several approximations, but results in a direct integration of the equations of motion and is capable of modeling the full laser damage process, from laser interaction to crater formation, allowing for direct experimental verification of the results with zero fitting parameters. This method was demonstrated here using a flat copper slab target, irradiated with an 800 nm, 60 fs laser pulse at normal incidence, which resulted in a damage threshold fluence of 0.15 J/cm$^2$, consistent with experiment. This simulation method however is flexible and is capable of modeling a wide variety of metals and dielectrics, even with non-trivial surface structures. Future studies are planned on expanding this method to more complicated targets.

This work was supported by the Air Force Office of Scientific Research grant no. FA9550-12-1-0454, and allocations of computing time from the Ohio Supercomputing Center.